# About the Protein Space Vastness[♣]

Jorge A. Vila

IMASL-CONICET, Universidad Nacional de San Luis, Ejército de Los Andes 950, 5700 San Luis, Argentina.

**Abstract**

An accurate estimation of the Protein Space size, in light of the factors that govern it, is a long-standing problem and of paramount importance in evolutionary biology, since it determines the nature of protein evolvability. A simple analysis will enable us to, firstly, reduce an unrealistic Protein Space size of ~$10^{130}$ sequences, for a 100-residues polypeptide chain, to ~$10^9$ functional proteins and, secondly, estimate a robust average-mutation rate per amino acid ($\xi$ ~1.23) and infer from it, in light of the protein marginal stability, that *only* a fraction of the sequence will be available at any one time for a functional protein to evolve. Although this result does not solve the Protein Space vastness problem, frames it in a more rational one.

At first glance, the universe of possible sequences populating the Protein Space (Smith, 1970), for a 100 residues polypeptide chain, should contain $20^{100}$ or ~$10^{130}$ sequences (Mandecki, 1998; Dryden *et al*. 2008; Romero & Arnold, 2009; Ivankov, 2017); where 20, the number of naturally occurring amino acids, represent the mutation rate per amino acid ($\xi$). In light of protein evolvability, a Protein Space (PS) of such size is absurdly huge. This problem is analog to the "Levinthal paradox" (Levinthal, 1969; Zwanzig *et al*. 1992; Finkelstein & Garbuzynskiy, 2013). According to this paradox, exploring the whole conformational space, in search of the native-state of a 100 residues polypeptide chain would require ~$10^{27}$ years (Zwanzig *et al*. 1992), *i.e.*, a time

---

[♣] I am honored to dedicate this essay to the memory of Harold A. Scheraga, Todd Professor of Chemistry, Emeritus at Cornell University. He achieved leadership in the world of science, and high respect among colleagues, as a result of his colossal experience in Experimental and Theoretical Chemistry, Physics and Mathematics, research in Protein Chemistry and, in particular, due to his tireless effort in the search of a possible solution to the, yet unsolved, Protein Folding problem. Harold Scheraga passed away at the age of 98 in Ithaca, NY, on August 1, 2020.

larger than the age of the universe, according to the Big Bang theory; yet proteins can fold in solution within microseconds to seconds. What do the Levinthal paradox and the Protein Space problem have in common? Certainly, it is not the size nor the time needed to explore it but our ignorance of the main forces that govern it. As far this work is concerned, the protein marginal stability is one of the forces limiting the vastness of the PS and, although it as being studied profusely in the last 50 years (Bloom *et al*., 2006; Wagner, 2005; DePristo *et al*., 2005; Zeldovich *et al*., 2007; Tokuriki *et al*, 2008; Romero & Arnold, 2009; Tokuriki & Tawfik, 2009, and reference therein) many questions remain to be answered. Among others, given a polypeptide sequence, which is a reasonable number of functional proteins in the Protein Space? For this reason, we start by reexamining the critical role of protein marginal stability on the estimation of the PS size; explicitly, the impact of this restraint on destabilizing mutations (Tokuriki & Tawfik, 2009; Martin & Vila, 2020). Then, after introducing simple approaches to estimate the PS size, we discuss the accuracy, in light of molecular evolution, of the main factors regulating such estimation. It is not in the spirit of this letter to discuss the navigation routes over this space by natural selection (Romero & Arnold, 2009; Tokuriki & Tawfik, 2009; Otwinowski, 2018) nor review existing proposals to estimate which fraction could have been explored since life began on Earth (Dryden *et al*., 2008). In other words, we are aimed to estimate for a protein all possible functional sequences in the 'Protein Space' in light of the proteins' marginal stability.

One of the central problems in the study of the evolvability of proteins is the influence of the 'protein stability' on the Protein Space size. As a first step in the analysis, a definition of 'protein stability' must be adopted because there are two ways to measure it. Specifically, by determining either the "denaturation" free-energy, $\Delta G^D$ (Koehl & Levitt, 2002) or the protein "marginal" stability (Privalov & Tsalkova, 1979; Hormoz, 2013; Vila, 2019; Martin & Vila, 2020). The latter but not the former definition is a more accurate way to study the PS size, and the reason for this assertion follows. In the PS model "…*two sequences are neighbors if one can be converted into another by a single amino-acid substitution* …" (Smith, 1970). This simple requirement enables us to use the $\Delta\Delta G^D$ free-energy changes between the wild-type and the mutated protein as a tool to estimate the feasibility of a single amino-acid substitution. These $\Delta\Delta G^D$ values reflect, primarily, changes in the native-state rather than in the unfolded-states (Zeldovich *et al*., 2007). Because of this, we consider that the 'marginal stability' is, in light of the PS model, a more accurate way to characterize the 'protein stability'. In this regard, we have been able to demonstrate, based on simple concepts from statistical thermodynamics, the Gershgorin theorem and a heuristic argument, the existence of an upper limit to the protein marginal stability (Vila, 2019).

In his seminal 1970 article, John Maynard Smith wrote: "*…if evolution by natural selection is to occur, functional proteins must form a continuous network which can be traversed by unit mutational steps without passing through nonfunctional intermediates…*" Implicit in this proposal is that any functional protein that pertains to the Protein Space must fulfill Anfinsen's dogma (Anfinsen, 1973); consequently, prion and IDP proteins will be excluded from our analysis. For the practical purpose, we depict the PS as a network of strings containing all functional proteins that nature devised. Each string contains ($m$) sequences, with a fixed number of residues ($n$), conforming a protein ensemble that fulfill Anfinsen's dogma and where each one of them can be converted into another by a single amino acid substitution. Nearest neighbor strings contain ($n+1$; $n-1$) residues. The string with the minima number of residues able to form a stable protein structure is assumed to contain $n = 16$ residues, *e.g.*, as the *β*-harping substructure of the immunoglobin binding domain of streptococcal protein *G*, because it form a stable native state in the absence of the rest of the protein (Cecchini, *et al.*, 2009). It is worth noting that metamorphic proteins, as protein *G* (Spichty, *et al.*, 2010), may appear in any string. However, this is not an obstacle for the analysis because this kind of proteins fulfill Anfinsen's dogma (Vila, 2020).

After clarifying some key definitions, the following question arises: given a string containing an arbitrary number of residues ($n$), which is the size of the corresponding Protein Space? Let us analyze it. As a simple and factual assumption, we envision that each amino acid in the sequence may be substituted by the one possessing the highest substitution rate in the Point Accepted Mutation (PAM1) matrix (Jones *et al.*, 1992; Gillespie, 1994). The adoption of this simple approach ($\xi = 2.0$) will enable us to predict quickly an order of magnitude for the PS size. Indeed, for a sequence of $n=100$ residue, there will be $m = 2^{100}$ or $\sim 10^{30}$ different sequences. However, this size for the Protein Space can be predicted by an even simplest approach. In fact, the modeling of a protein as a collection of hydrophobic or polar (HP) beads (Lipman & Wilbur, 1991; Bornberg-Bauer, 1997) also enables to predict a PS size of $\sim 10^{30}$. But, there are two observations to note about the accuracy of this prediction. Firstly, the mutation rate under the HP modeling is unsound in light of molecular evolution. Secondly, the protein representation under the HP model foresees a challenge to Anfinsen's dogma, *e.g.*, has been shown that a two-letter code is insufficient to give an energy landscape like that of a wild type protein (Wolynes, 1997). Let us briefly discuss the relevance of each of these objections. Under the HP modeling a single-point mutation is simply a replacement of a hydrophilic (H) residue by a polar (P) one, and vice versa ($\xi = 2.0$). This approach ignores key mutations seen in nature, *e.g*., mutation between polar residues in hemoglobin (Perutz, 1983). Moreover, whether or not the protein HP model challenges the Anfinsen dogma it is not a minor problem. On the contrary, this is a serious problem because the marginal stability of the proteins is one of the most important factors restricting their evolvability and its existence is a consequence of Anfinsen's dogma (Vila, 2019). In this regard, should be noted that the protein marginal stability is a universal property, that includes biomolecules and macromolecules complex (Martin & Vila, 2020), and their origin can be thoroughly understood from a purely physical point of view (Vila, 2019) or, alternatively, from specific evolutionary arguments (Wilson *et al*., 2020).

Whatever the above-adopted model to estimate the PS size is the impact of the protein marginal stability implies that its actual size shall be significantly smaller than foretold ($\sim 10^{30}$) because most of the single-site mutations in real proteins are destabilizing (Zeldovich *et al*., 2007; Tokuriki & Tawfik, 2009). Moreover, it is well documented, from evolutionary changes observed in cytochromes *c* of various species, that many sites are invariant to mutations (Margoliash & Smith, 1965). In other words, substitutions at specific sites in the protein sequence may not ever occur and the reason is the following. Proteins for which single-site substitutions lead to a free-energy change ($\Delta\Delta G_D$) larger than the marginal stability upper bound limit (~7.4 Kcal/mol) will never fold or function (Martin & Vila, 2020). This destabilization threshold value agrees with observations made on the green fluorescent protein from Aequorea victoria (avGFP), *viz*., "…*The average fluorescence of single mutants of avGFP as a function of the predicted protein destabilization, ΔΔG, reveals a threshold around 7–9 kcal mol$^{-1}$* …" (Sarkisyan *et al*., 2016). In particular, is worth noting that a sigmoid-like fitness function obtained by an independent neural-network analysis predicts a ~ 60% drop on the log-fluorescence if $\Delta\Delta G > \sim 7.5$ kcal/mol (see Fig. 4 of Sarkisyan *et al*., 2016). Putting all together, the prediction of a reasonable Protein Space size requires a model reappraisal in light of the proteins' marginal stability.

Evidences have been presented showing that "…*more stable proteins are more evolvable because they are better able to tolerate functionally beneficial but destabilizing mutations*…" Bloom *et al*., (2005). Our concern about this proposal comes from the fact that the upper bound free-energy gap between the native-folds and the misfolded ones is very small; to be specific, equivalent to break, at most, ~5 hydrogen-bonds! (Martin & Vila, 2020). In other words, protein evolvability need to overcome this drawback by using a very efficient mechanism. A possible solution to this problem has already been anticipated by Kimura (1968) and highlight by Wagner (2005) and Bloom *et al*. (2005) by suggesting that neutral mutations may play a critical role in the transition from one structure to another in the Protein Space, *e.g*., by counterbalance the effects of destabilizing mutations although beneficial from the functional point of view. From this perspective and considering that each amino acid is coded by a nucleotide triplet, an *upper bound limit* for the PS size can be estimated. Let do it by assuming, firstly, $n=100$ as a trial length, although the analysis would be valid for sequences of any length and, secondly, that: (*i*) each nucleotide pair replacement entails an amino acid substitution; (*ii*) each nucleotide pair replacement occurs, after removing synonymous mutations, every ~2 years (Kimura, 1968); and (*iii*) almost all mutations will be neutral (Kimura, 1968) or nearly neutral (Otha, 1973). If the starting point is a functional protein, adoption of these rules will assure that the Protein Space will be a "…*continuous network which can be traversed by unit mutational steps without passing through nonfunctional intermediates*…" Smith (1970). Consequently, if life began on earth around a billion years ago (Schopf, 2006) then the Protein Space should contain ~$10^9$ functional sequences. Therefore, the average mutation rate per amino acid will be $\xi$ ~1.23. Before we continue, let us analyze some nucleotide replacement alternatives to the one proposed above, in section (*ii*). For example, we could have considered one nucleotide pair replacement per day or ~$10^{14}$ per second rather than one every ~2 years (Kimura, 1968). These alternatives will lead to Protein Space sizes containing ~$10^{11}$ ($\xi$ ~1.30) or ~$10^{30}$ ($\xi$ ~2.0) functional sequences, respectively. These Protein Space sizes seem to be reasonable although the nucleotide pair replacement rates are not. This simple example illustrates that a reliable estimation of the Protein Space size, in light of molecular evolution, is not just a combinatorial problem. Indeed, time is an essential variable to find out a reasonable answer, *i.e*., as in the search of solutions for the Levinthal paradox (Zwanzig *et al*. 1992). So, an average mutation rate per amino acid of $\xi$ ~1.23 is equivalent to think that *only* a fraction of the protein sequence is variable at any one time, *e.g*., 30% with $\xi = 2.0$; 13% with $\xi = 4.0$; and 5% with $\xi = 8.0$, while mutations on the remaining of the sequence would lead to conformations that do not fold; in other words, will be nonfunctional. That each functional protein in the Protein Space, independent on the fold-class or sequence, could tolerate only a fraction of their sequence to be variable at any one time, is consistent with the conjecture that protein marginal stability's is the main force that confines the Protein Space size; basically, because the marginal stability sets up a physical limit to the amount and type of mutations that a protein can admit while still folding into the native structure (Martin & Vila, 2020). Hence, it is not surprising to read that ~25% of the single-site mutations on the green fluorescent protein (from avGFP) have not deleterious effects on fluorescence (Sarkisyan *et al*., 2016), or that many sites are invariant to mutations in cytochrome *c* (Margoliash & Smith, 1965).

There are at least two hidden assumptions in this proposal, just to mention a few. Firstly, the estimated Protein Space size (~$10^9$) represents *all* functional forms compatible with one functional protein that existed since life started on Earth. This does not necessarily mean that all the functional proteins come from a single starting point. Indeed, as noted by John Maynard Smith 50 years ago, is possible that "…*there are two or more distinct networks, or that there is one network with multiple starting points*…" Secondly, the use of the beginning of the life on Earth as a reference point does not mean, nor imply either, that evolution was able to explore the whole Protein Space or how evolution occurred during such a period of time, *i.e*., either by the natural (Darwinian) selection of favorable mutations or by random fixation of neutral or nearly neutral mutants.

Overall, the existence of intermedia steps during the molecular evolution, like those brought by neutral or nearly neutral mutations, enable us, firstly, to conjecture a conceivable Protein Space size of ~$10^9$ for a starting 100-residues functional protein and, secondly to estimate a robust average-mutation rate per amino acid ($\xi$ ~1.23), *i.e*., independent of the protein fold-class, or sequence, and infer, in light of the protein marginal stability, that *only* a fraction of the sequence will be available at any one time for a functional protein evolve.

## Acknowledgments

I would like to thank to Laura Mascotti, Walter Lapadula and Maximiliano Juri Ayub for reading the manuscript and making valuable comments and suggestions. The author acknowledges financial support from the IMASL-CONICET (PIP-0087) and ANPCyT (PICT-0767; and PICT-2212), Argentina.